\documentclass[preprint,1p]{elsarticle}

\usepackage{lineno,hyperref}
\usepackage{wrapfig}
\modulolinenumbers[5]

\journal{Nano Energy}

\begin{document}

\begin{frontmatter}

\title{An iterative machine learning approach for discovering unexpected thermal conductivity enhancement in aperiodic superlattices}


\author[Purdue]{Prabudhya Roy Chowdhury}

\author[Purdue]{Xiulin Ruan\corref{corauth1}}

\address[Purdue]{School of Mechanical Engineering and the Birck Nanotechnology Center, Purdue University, West Lafayette, Indiana 47907-2088, USA}

\cortext[corauth1]{Corresponding author:\\
\textit{Email addresses:} ruan@purdue.edu (Xiulin Ruan)}

\begin{abstract}
While machine learning (ML) has shown increasing effectiveness in optimizing materials properties under known physics, its application in challenging conventional wisdom and discovering new physics still remains challenging due to its interpolative nature. In this work, we demonstrate the potential of using ML for such applications by implementing an adaptive ML-accelerated search process that can discover unexpected lattice thermal conductivity ($\kappa_l$) enhancement instead of reduction in aperiodic superlattices (SLs) as compared to periodic superlattices. We use non-equilibrium molecular dynamics (NEMD) simulations for high-fidelity calculations of $\kappa_l$ for a small fraction of SLs in the search space, along with a convolutional neural network (CNN) which can rapidly predict $\kappa_l$ for a large number of structures. To ensure accurate prediction by the CNN for the target unknown structures, we iteratively identify aperiodic SLs containing structural features which lead to locally enhanced thermal transport, and include them as additional training data for the CNN in each iteration. As a result, our CNN can accurately predict the high $\kappa_l$ of aperiodic SLs that are absent from the initial training dataset, which allows us to identify the previously unseen exceptional structures. The identified RML structures exhibit increased coherent phonon contribution to thermal conductivity owing to the presence of closely spaced interfaces. Our work describes a general purpose machine learning approach for identifying low-probability-of-occurrence exceptional solutions within an extremely large subspace and discovering the underlying physics.

\end{abstract}


\end{frontmatter}


\section{Introduction}

The demand for efficient energy systems and high-performance electronic devices has created the challenging requirement to rapidly identify new materials and design nanostructures with extreme transport properties. As the limitations of traditional intuition-driven trial-and-error search methods become more prominent, machine learning (ML) and data informatics have emerged as powerful tools for solving these design and optimization problems. In thermal transport, ML methods have found success in predicting material properties and accelerating design of nanostructures with target thermal transport \cite{ermis2007heat, esfe2014thermal, ju2017designing, yang2018machine, wei2018predicting, rong2019predicting, chowdhury2020machine, chakraborty2020quenching, wei2020machine}. However, the applications of ML to solve thermal engineering problems till date have been limited to finding optimal solutions which confirm previously well understood phonon transport theory. ML has not yet been used to explore and discover exceptional solutions which can help us uncover new facets of phonon transport theory and guide the design of novel nanostructures. This can be attributed to the ``interpolative'' nature of traditional ML algorithms which allows for accurate prediction and exploration within the subspace spanned by known data points (and, therefore, known physics), but fails for excursions outside the training dataset. Consequently, suitable adaptations are needed to use ML methods in the identification of materials or nanostructures showing exceptional physical properties.

In this work, we demonstrate the potential of an adaptive machine learning approach to identify unexpected thermal transport behavior in aperiodic superlattices. Binary superlattices (SLs), composed of periodically alternating layers of two materials, have received widespread attention in the recent decades due to their lower lattice thermal conductivity ($\kappa_l$) compared to the constituent materials \cite{lee1997thermal, simkin2000minimum, chen1998thermal, venkatasubramanian2000lattice}, which makes them greatly attractive for applications such as thermoelectric devices \cite{venkatasubramanian2001thin, bottner2006aspects, chowdhury2009chip}. Recent studies have shown that randomizing the constituent layer thicknesses in periodic SLs further reduces $\kappa_l$, even below the random alloy limit \cite{frachioni2012simulated, PhysRevB.90.165406, wang2015optimization, mu2015ultra, chowdhury2020machine, frieling2016molecular, doi:10.1080/15567265.2015.1102186, chakraborty2017ultralow}. In the resulting aperiodic superlattices or random multilayers (RMLs), destructive interference of coherent phonons due to reflections at the randomly spaced interfaces can cause Anderson localization, thereby limiting thermal transport by these long wavelength phonon modes \cite{PhysRevB.90.165406, juntunen2019anderson}. Additionally, ML methods such as Bayesian optimization \cite{ju2017designing} and genetic algorithms (GA) \cite{chowdhury2020machine} have allowed efficient identification of RML structures with ultra-low thermal conductivities at a fraction of the computational cost associated with exhaustively searching the prohibitively large set of candidate structures. However, it has not yet been elucidated whether certain random distributions of SL layer thicknesses can actually lead to higher $\kappa_l$ than the periodic SLs. Interestingly, in a recent study, Wei \textit{et al.} \cite{wei2020genetic} used a GA-based search process to identify two-dimensional graphene nanomeshes with disordered pore configurations showing enhanced $\kappa_l$ than nanomeshes with uniformly spaced pores. Their results challenged the previous understanding that randomness in pore spacings leads to lower $\kappa_l$ in these systems \cite{feng2016ultra,hu2018randomness}. Although heuristic search techniques such as GAs are known to be ``extrapolative'' due to their ability to explore the design space outside the initial known dataset, they are still computationally expensive due to the requirement for a predictor step at each iteration of exploration. Nonetheless, such demonstrations drive the search to find exceptions for other well understood systems such as SLs and RMLs, where such solutions constitute a very low fraction of the design space. Therefore, we look to find a systematic approach which can utilize the advantages of ML while enabling an extrapolative approach to efficiently identify these low-probability-of-occurrence novel solutions.

Here, we identify RML structures with unexpectedly higher $\kappa_l$ than corresponding SLs with same total length and average period. To accelerate the search over the prohibitively large design space, a convolutional neural network (CNN)-based prediction method is used for obtaining the $\kappa_l$ of the candidate structures. An iterative approach is employed for generating a representative training dataset that enables the CNN to accurately predict the high $\kappa_l$ of the target RML structures that are absent from the initial dataset. Finally, the identified non-intuitive RML structures are used to gain insight into the heat transport mechanisms leading to higher $\kappa_l$ and its correlation with RML structural features.

\section{Simulation methods}

\subsection{Simulation materials and multilayered structures}

We perform our calculations on the model Si/Ge system to search for high $\kappa_l$ RML structures. This system has been extensively investigated in literature, given the wide application of these semiconductor materials as multilayer systems \cite{PhysRevB.86.235304, zhang2007simulation, gordiz2016phonon, feng2019unexpected, PhysRevB.79.075316, sun2010molecular, PhysRevB.85.195302} and the simplicity of performing molecular dynamics simulations using interatomic potential parameters. The SLs and RMLs are constructed by stacking the diamond cubic unit cells (UCs) of each material along the [100] direction. Two different lengths of SL and RML structures are studied in this work: a shorter 20 UC (11 nm) system and a longer 40 UC (22 nm) system. Periodic boundary conditions are maintained in the other two directions, so that our system results in a superlattice thin film. A $6 \times 6$ UC cross-section is used, which is sufficient to provide converged $\kappa_l$ values. The smallest layer thickness allowed along the cross-plane heat transport direction is set to be 1 UC, and only RMLs with equal number of Si and Ge layers are studied to ensure meaningful comparison of $\kappa_l$ among all structures. Additionally, the first and last UCs along the RML length are constrained to be Si and Ge respectively, to prevent extra interfaces with the heat reservoirs. With these constraints imposed, the number of possible RML structures is found to be 48620 for the 20 UC system and 35345263800 for the 40 UC system.

\subsection{Non-equilibrium molecular dynamics simulations}

\begin{figure}[tb]
\centering
\includegraphics[scale=0.5]{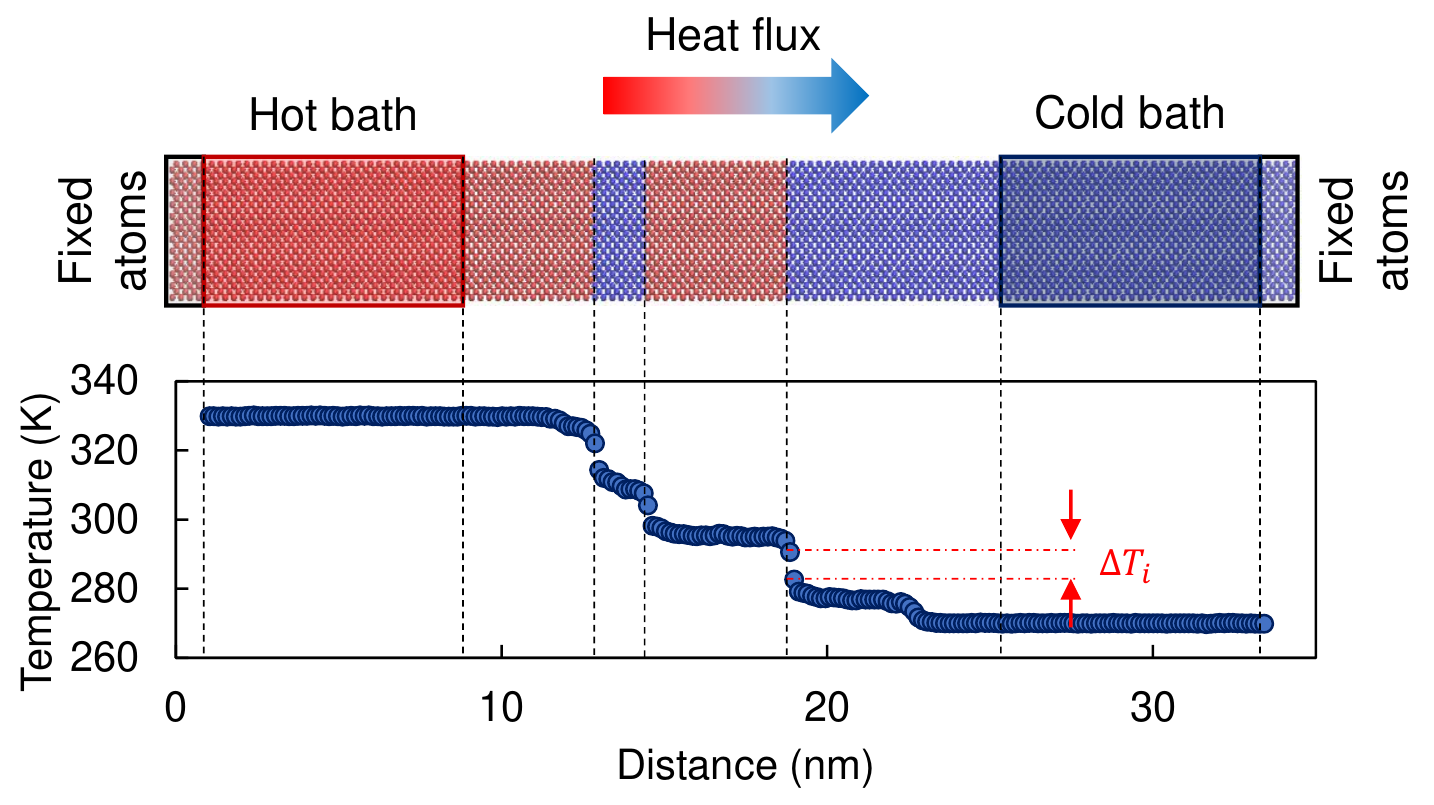}
\caption{\label{fig:NEMD} Schematic of the NEMD simulation setup showing the multilayer nanostructure (SL or RML) sandwiched between two thermal baths. A layer of atoms is fixed at each end to impose fixed boundary conditions. The corresponding temperature profile is also shown.}
\end{figure}

Thermal conductivity calculations for the multilayered nanostructures are performed using non-equilibrium molecular dynamics simulations with the LAMMPS package~\cite{plimpton1993fast}. The interatomic interactions are described using the three-body Tersoff potential~\cite{PhysRevB.39.5566, PhysRevB.41.3248.2}, which is commonly used to study vibrational properties of the Si/Ge system. The unequal equilibrium lattice constants of Si and Ge in these potential descriptions leads to a symmetric cross-sectional strain in the system, which can cause large oscillations at the interface regions~\cite{feng2019unexpected}. To eliminate this strain, the lattice constant of Ge is artificially set to be equal to that of Si within the interatomic Tersoff potential parameters. The thermal conductivity of the nanostructures is calculated at a temperature of 300 K. A timestep of $0.5$ fs is used to integrate the equations of motion, which is sufficient to resolve the highest frequency of phonon vibrations in either material.

A schematic of the NEMD simulation domain for direct calculation of thermal conductivity is shown in Fig.~\ref{fig:NEMD}. Two bulk material regions consisting of 20 UCs of Si and Ge are attached to either side of the SL or RML to act as thermal reservoirs. Initially, the entire system is relaxed for 500 ps at 300 K, under a constant pressure and temperature ensemble (NPT) with periodic boundary conditions applied to all three directions. Following this, another 250 ps of equilibration under fixed volume and energy (NVE) is performed. Subsequently, non-equilibrium conditions are applied by thermostatting the Si and Ge bulk regions on either side at 330 K and 270 K respectively, using Langevin thermostats. Two UCs of atoms at each end of the system are also kept fixed to mimic fixed boundary conditions along the heat transport direction. The system is allowed to reach steady state under this imposed temperature gradient over a period of 500 ps. Following this, the temperatures at equal intervals along the cross-plane direction are obtained by from the velocities of atoms in one-dimensional bins. The temperature and heat flux data is collected and averaged over a period of 4 ns. The cross-plane lattice thermal conductivity ($\kappa_l$) is then calculated as

\begin{eqnarray}
\kappa_l = \frac{q''}{\Delta T/L}
\label{eq:nemd}
\end{eqnarray}

Here, $q''$ is the steady state heat flux and $L$ is the length of the SL or RML along the heat transport direction. The thermal boundary resistance at each interface of the system ($R_i$) can also be calculated from the temperature drop across the interface ($\Delta T_i$) as

\begin{eqnarray}
R_i = \frac{\Delta T_i}{q''}
\label{eq:tbr}
\end{eqnarray}

\subsection{Convolutional neural network-based prediction of thermal conductivity}

\begin{figure*}[tb]
\centering
\includegraphics[scale=0.45]{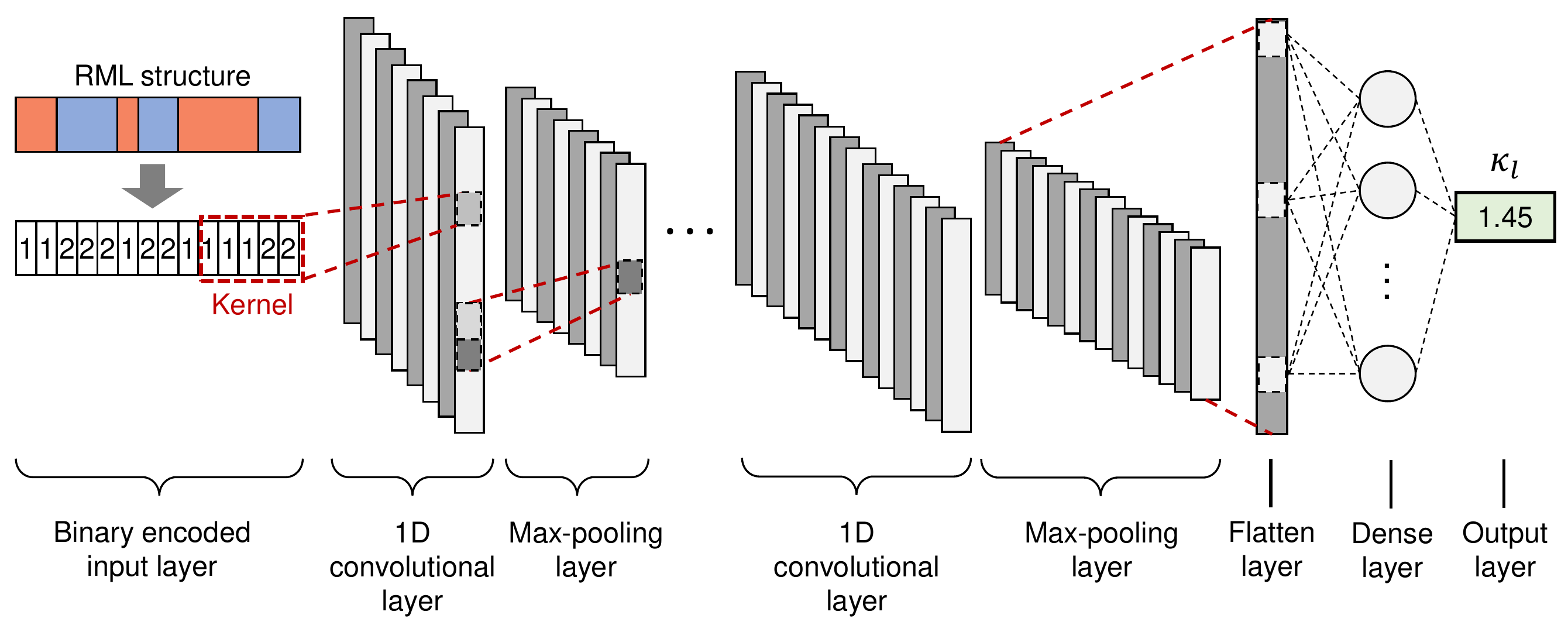}
\caption{\label{fig:NNarc} Schematic of the convolutional neural network architecture. The SL or RML structure is encoded as a binary array and used as the input layer, while a single output node provides the predicted thermal conductivity.}
\end{figure*}

While NEMD simulations can provide accurate values of thermal conductivity of the superlattice structures using a simple calculation framework, they are computationally expensive when more than hundreds of simulations need to be performed for a particular system. As a result, exhaustive searches using MD simulations over design spaces as large as the current problem become impractical. In order to accelerate the calculation of thermal conductivity of RMLs and perform a rapid screening of a large number of candidate solutions, we use a neural network prediction tool which can predict the thermal conductivity from the RML structure as input. Neural networks (NNs) have emerged as a powerful tool for regression and classification problems due to their ability to fit complex multifunctional datasets without the need for encoded sets of rules which may introduce human bias. Recently, convolutional neural networks (CNNs) have been successfully used in predicting material properties including $\kappa_l$ from input structure data \cite{wei2018predicting,rong2019predicting}, particularly due to their feature detection and translational invariance characteristics. 

The architecture of the CNN used in this work is shown in Fig.~\ref{fig:NNarc}. The input layer to the CNN is an N-bit array, corresponding to the number of UCs in the RML structure (20 or 40). Each bit can take a value of 1 or 2 depending on whether the corresponding UC at that location along the superlattice length consists of Si or Ge atoms respectively. This is followed by one or more one-dimensional convolutional layers, each of which consist of several kernels or filters to extract the relevant features from the input array by striding over the length of the input. Here, we use convolutional layers consisting of $44-50$ filters with filter lengths of $5-9$ bits, a stride length of 1 and no-padding. A max pooling layer is used after every two convolutional layers, which causes down-sampling of the identified features and incorporates translational invariance in the feature maps. After multiple convolutional layers, we use a flatten layer to reduce the dimensionality of the features. Finally, a fully connected or dense layer consisting of 100 nodes is used to combine the identified features into a single output thermal conductivity value. Non-linearity is accounted for within the CNN by using a Rectified Linear Unit (ReLU) as the activation function throughout the network. For the 20 UC RML system, we use a CNN consisting of 2 convolutional layers, 1 max-pooling layer and 1 fully connected layer. On the other hand, for the 40 UC RML system where the number of input parameters is much larger, we switch to a CNN architecture consisting of 4 convolutional layers with 1 max-pooling layer after every 2 convolutional layers, and 1 fully connected layer as before.

The weights of the different layers are initiated randomly and need to be fit to the training data provided to the network. This is done by calculating a loss function over the entire training set and back-propagating the errors over the various layers of the network to minimize the loss. The loss function used to train our CNN is chosen to be the mean absolute percentage error ($MAPE$), given by

\begin{eqnarray}
MAPE = \frac{1}{N}\sum_{i=1}^{N} |\frac{y_i - \bar{y}_i}{\bar{y}_i}|\times100\%
\label{eq:tbr}
\end{eqnarray}

Here, $N$ is the number of training data points provided to the network, $y_i$ is the predicted output and $\bar{y}_i$ is the target output. Apart from the loss function, the root mean square error (RMSE) is also used a metric to evaluate the performance of the network. We note that these metrics are most commonly associated with regression problems, instead of others such as accuracy which are convenient for classification tasks. The training of the network by back-propagation of errors is performed using the Adamax algorithm~\cite{kingma2014adam} and the fitting is performed over $300-500$ epochs within which sufficient convergence of the loss function is observed. Overfitting of the data by the CNN, which is common occurence in neural network training, is avoided using early stoppage of the fitting process if the testing loss is found to become constant or increase. Once the CNN is trained, it can be used to predict the thermal conductivities of the entire dataset of RML structures within several seconds, thereby making an exhaustive search possible. The most time intensive part of the neural network based prediction process is generation of a representative training dataset containing a reasonable number of data points, which can be done systematically to a great advantage as explained in the subsequent section.

\subsection{Iterative approach for generating the training dataset during search}

An important characteristic of neural networks is their ``interpolative'' nature, \textit{i.e.} they cannot generally be expected to extrapolate to unknown points outside the region spanned by the training dataset. This is problematic for our current objective, where the CNN is required to accurately predict thermal conductivities of high $\kappa_l$ exceptional RMLs which are absent from the initial training dataset. To resolve this, we utilize the ability of CNNs to extract spatial features contributing to locally enhanced thermal transport. Although the training dataset is composed of RML structures with low to moderate $\kappa_l$, many of these structures contain spatial features that lead to locally enhanced thermal transport, such as large bulk regions or short regions of periodic interfaces. By forming feature-property maps from these structural features, the CNN is able to assimilate them and accurately predict the high $\kappa_l$ of RMLs containing combinations of these favorable features.

\begin{figure}[t]
\centering
\includegraphics[scale=0.45]{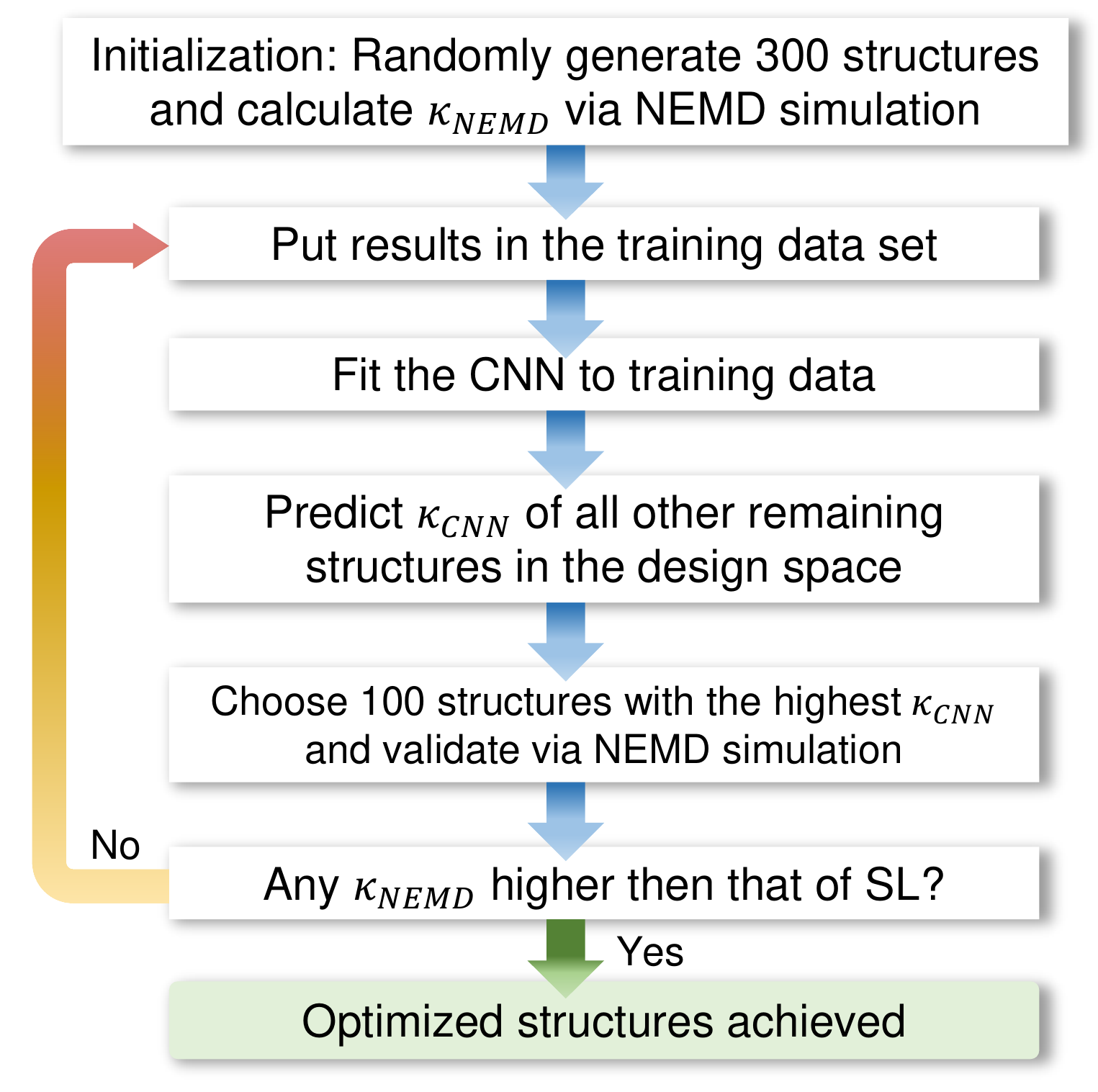}
\caption{\label{fig:training_flow} Schematic of the iterative search algorithm used to discover unexpected thermal conductivity ($\kappa_l$) enhancement in aperiodic superlattice systems.}
\end{figure}

On the other hand, randomly sampling the design space does not automatically ensure inclusion of RML structures showing enhanced local thermal transport characteristics within the dataset. In order to overcome this challenge, we adopt an iterative approach to dynamically generate our training dataset comprising RMLs with moderate to high thermal conductivities while performing the accelerated search. In the initial step, the CNN is trained on a dataset of the 300 randomly generated RMLs. The trained network is then used to predict the thermal conductivities ($\kappa_{CNN}$) of all structures in the search space. Next, we select 100 RML structures predicted by the CNN to have the highest thermal conductivities and perform NEMD calculations of thermal conductivity ($\kappa_{NEMD}$) to validate the CNN predicted values. If any of these 100 RML structures identified in the search show a higher $\kappa_{NEMD}$ than the corresponding SL, the search is stopped with successful identification of the exceptional RML structures. Otherwise, these RML structures are included in the training data with their $\kappa_{NEMD}$ values, and the CNN is retrained to fit the augmented data set. Subsequently, the thermal conductivities of all structures are again predicted (with potentially higher accuracy) and the algorithm is progressed in this manner. Figure~\ref{fig:training_flow} shows the complete work flow of the search algorithm followed in our work. In the initial iteration, the $\kappa_{CNN}$ values are not expected to be accurate over the entire search space, given the relatively small size of the training dataset and the absence of representative features. However, the accuracy of prediction improves as the size of the training dataset increases with each successive iteration and RMLs with high $\kappa_l$ constitute a greater fraction of the training data.

\section{Results and discussions}

\subsection{Manual search for higher thermal conductivity RMLs}

\begin{figure}[t]
\centering
\includegraphics[scale=0.65]{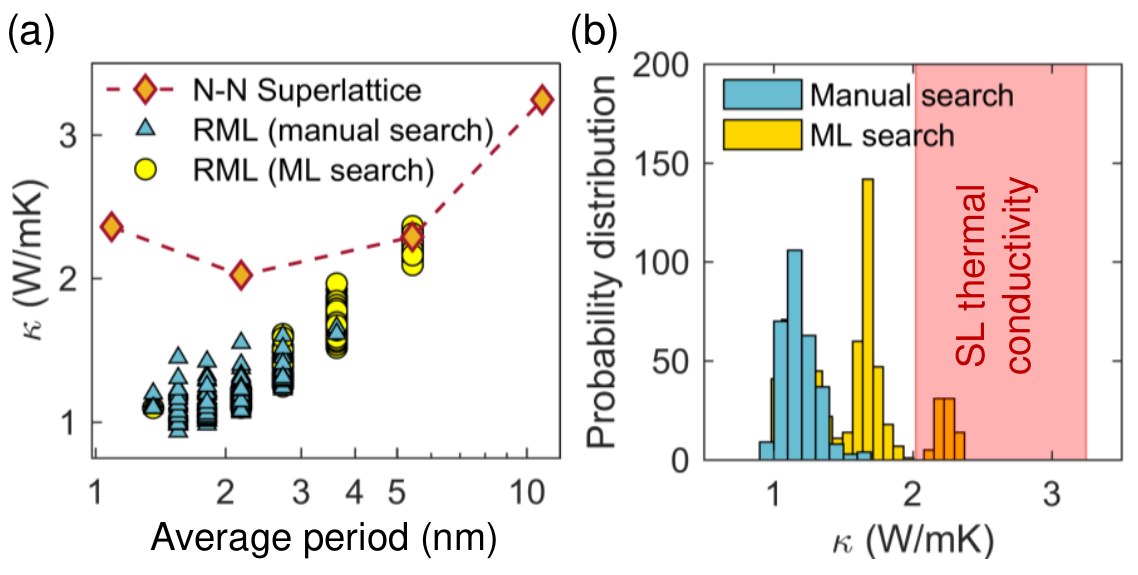}
\caption{\label{fig:kperdist} (a) Variation of $\kappa_l$ with average period at 300 K for RML structures generated during the manual random search (triangles) and the machine learning accelerated search (circles).The thermal conductivities of the reference $N-N$ superlattices are indicated by the diamonds. (b) Probability distributions of thermal conductivities (W/mK) of the RML structures generated by a manual search (blue bars) and the ML search (yellow bars). The region spanned by the thermal conductivities of the $N-N$ SLs is shaded in red.}
\end{figure}

First, we search for 20 UC (10 nm) RMLs showing enhanced $\kappa_l$ from the corresponding SL structures. The thermal conductivities of the 20 UC $N-N$ SL system are calculated, where $N$ is the number of unit cells of Si or Ge in one period of the SL. To ensure an integral number of periods within the fixed total length of 20 UCs, $N$ can take values of 1,2,5 and 10 only. The thermal conductivities obtained using NEMD simulations are shown in Fig.~\ref{fig:kperdist} (a), where a minimum of 2 W/mK is obtained at an SL period of $\sim 2.2$ nm. This characteristic variation of $\kappa_l$ with SL period has been predicted theoretically \cite{simkin2000minimum, venkatasubramanian2000lattice, chen2005minimum, PhysRevB.66.024301} and recently observed experimentally \cite{caylor2005enhanced, luckyanova2012coherent, ravichandran2014crossover, PhysRevB.93.045311}, and is commonly understood to be the result of the transition from coherent phonon to incoherent phonon dominated transport regimes. Phonons travelling along the cross-plane direction of SLs with large periods can exhibit particle-like behavior when anharmonic phonon-phonon scattering causes them to lose phase information before encountering an interface. On the other hand, multiple phase-preserved reflections at closely spaced periodic interfaces can lead to the formation of coherent phonon modes showing wave-like phonon transport characteristics. At periods greater than $2.2$ nm, the interface density is small enough to ensure a low coherent phonon contribution. As a result, the reduction in incoherent phonon scattering by the SL interfaces leads to a greater thermal conductivity at higher periods. In contrast, when the SL period is below $2.2$ nm, a significant portion of the thermal transport is contributed by the coherent phonon modes, which are no longer scattered by the closely spaced interfaces. In this regime, the increase of thermal conductivity at lower periods has been attributed to effects such as weaker band flattening and increased group velocities.

We then attempt the traditional \textit{intuition-guided} search process to identify possible RMLs showing $\kappa_l$ enhancement due to aperiodicity. Due to the absence of any previous evidence supporting the existence of enhanced $\kappa_l$ RML structures, no guidance is available to narrow down the search to a computationally tractable subset of the design space. In this case, a random search can be considered to be one of the best possible search methods available. To perform the manual search, we randomly choose 300 candidate RML structures from the design space and calculate the thermal conductivities using NEMD simulations. The results of these calculations are compared with the SL thermal conductivities in Fig.~\ref{fig:kperdist} (a). We find, as expected, that all of the 300 randomly generated RMLs have significantly lower thermal conductivities than the corresponding SLs. This shows the evident need for an alternative systematic and efficient way to perform the search and motivates the use of machine learning for such tasks. We also calculated the histogram of thermal conductivity values for the 300 randomly generated RML structures as shown in Fig.~\ref{fig:kperdist} (b). It can be observed that the majority of RMLs have low thermal conductivities compared to the $N-N$ SLs. Consequently, the structural features leading to locally enhanced thermal transport are underrepresented in the training dataset. This leads us to follow an adaptive approach of generating training data whereby we can discover high $\kappa_l$ RMLs in the search space and integrate them within our training dataset.

\subsection{Machine learning accelerated search for higher thermal conductivity RMLs}

Due to the impracticality of using a random search to identify high thermal conductivity RMLs, we perform a machine learning accelerated search in which a convolutional neural network (CNN) is used to replace the time consuming NEMD simulations as a rapid thermal conductivity prediction tool. Since the time taken by the trained CNN to predict the $\kappa_l$ of each RML structure is extremely small, the entire design space can be evaluated within several seconds. As a result, an exhaustive search can be performed over the entire RML design space using the $\kappa_l$ values predicted by the CNN. 

\begin{figure}[ht]
\centering
\includegraphics[scale=0.45]{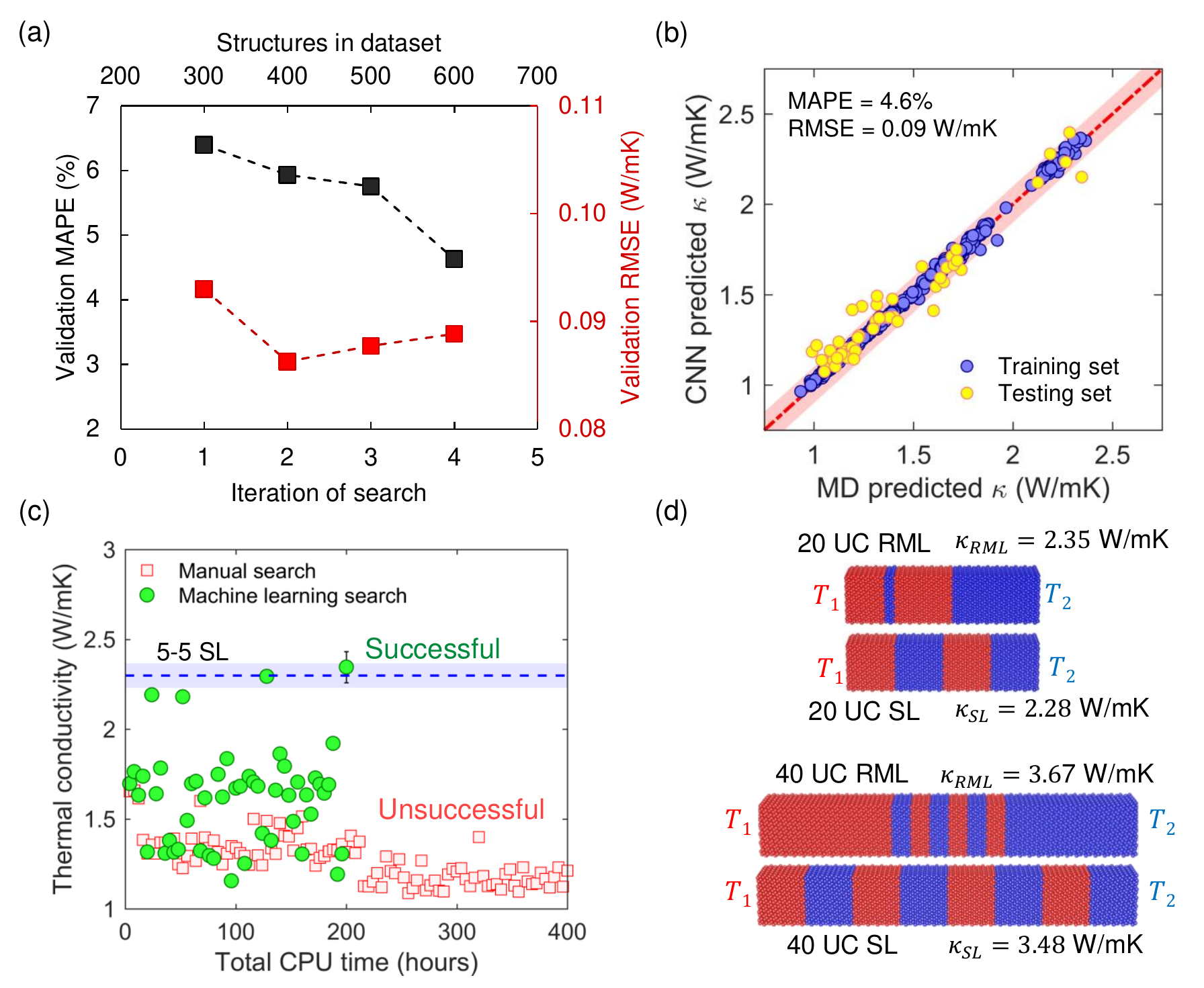}
\caption{\label{fig:NNresults} (a) Variation of testing MAPE (black squares, left axis) and RMSE (red squares, right axis) with each iteration of the iterative search process for the 20 UC RML system. The top axis indicates the size of the dataset on which the CNN is trained in that iteration of the search (b) Comparison of CNN predicted $\kappa_l$ and NEMD calculated $\kappa_l$ (true value) for the dataset of RML structures. The shaded area represents a $\pm 0.1$ W/mK bound from $y=x$. (c) Thermal conductivities of 20 UC RMLs sampled by the random search (squares) and the CNN accelerated search (circles) with total computational time spent. The dashed line represents the $\kappa_l$ of the 5-5 SL structure with error bounds. (d) The 20 UC and 40 UC RML structures with higher thermal conductivities than the corresponding SLs which were identified by the CNN accelerated search.}
\end{figure}

We first evaluate the performance of the CNN in predicting $\kappa_l$ of the 20 UC RML system. Figure~\ref{fig:NNresults} (a) shows the variation of the MAPE and the RMSE with each iteration of the search process, when evaluated on a testing set of unknown RML structures not introduced to the CNN during training. We observe that the CNN is able to predict thermal conductivities with a very low MAPE varying from $4.6-6.4\%$, or an average RMSE of $0.09$ W/mK. The MAPE generally decreases with each progressing iteration of the search due to the addition of more RML structures to the training dataset which increases the representative set of features. The comparison between the predicted ($\kappa_{CNN}$) and ``true'' values ($\kappa_{NEMD}$) is shown by the parity plot in Fig.~\ref{fig:NNresults} (b) after training the CNN on data from 600 RML structures. It is seen that the CNN can provide accurate predictions over a wide range of thermal conductivities from $1-2.5$ W/mK, thus demonstrating the capability of the CNN to extract suitable spatial features governing low and high $\kappa_l$. The progress of the ML enabled search for 20 UC RMLs with enhanced $\kappa$ are shown in Fig.~\ref{fig:NNresults} (c), in comparison to a manually performed random search. We only compare RML structures against the corresponding SL having the same average period. As a result, the contribution of interface scattering of incoherent phonons to the thermal transport is the same in the compared multilayer structures, and any difference in $\kappa_l$ is purely the result of coherent phonon transport. We find that our ML-based search process is able to identify RML structures with higher $\kappa$ than the corresponding SL within two iterations of the search utilizing 200 CPU hours. In contrast, the manual random search returns far lower $\kappa_l$ than periodic SLs even after double the simulation hours spent. The thermal conductivities of the RMLs scanned by the ML search process are plotted with respect to average period in Fig.~\ref{fig:kperdist} (a). By searching through RML structures with different average periods, the $\kappa_l$ of RML structures are found to exceed the superlattice $\kappa_l$ at a relatively higher average period of $5.4$ nm, corresponding to the $5-5$ SL. The identified RML $\kappa$ of 2.36 W/mK is found to be higher than the SL $\kappa$ of 2.28 W/mK by $3.5\%$, which is above the statistical uncertainty as confirmed by averaging these values over multiple independent runs. The structures of the $5-5$ SL and the RML showing enhanced $\kappa_l$ are shown in Fig.~\ref{fig:NNresults} (d).

We also perform a similar search for a larger RML system with a total length of 40 UCs. Since the number of possible RML structures for this system is several orders of magnitude larger than the 20 UC system, we limit our search to a tractable subset of the design space by using the knowledge gained from the results of the search on the 20 UC RML system. In particular, only RMLs with the relatively larger average period of $5.4$ nm, corresponding to perturbations of the $5-5$ SL, are sampled. With this constraint, the reduced design space consists of 938961 RML structures which can be efficiently handled by our ML search framework. Similar to the previous search process, the CNN accelerated search method can successfully identify an RML structure with higher $\kappa$ than the corresponding SL within validation of 612 RMLs which constitute less than 0.1\% of the design space. The identified structure, shown in Fig.~\ref{fig:NNresults} (d), has a $\kappa_l$ exceeding that of the SL by $5.5\%$ which is also confirmed by averaging over multiple runs. Interestingly, the 40 UC RML structure identified by our search is found to be a composite SL which can be created by combination of the single interface structure and the shorter period $2-2$ SL. As a result, the structure has the features of a local periodicity which enhances thermal transport despite having a globally random layer thickness distribution.

\subsection{Contribution of interfacial resistance towards $\kappa_l$ enhancement}

\begin{figure}[t]
\centering
\includegraphics[scale=0.7]{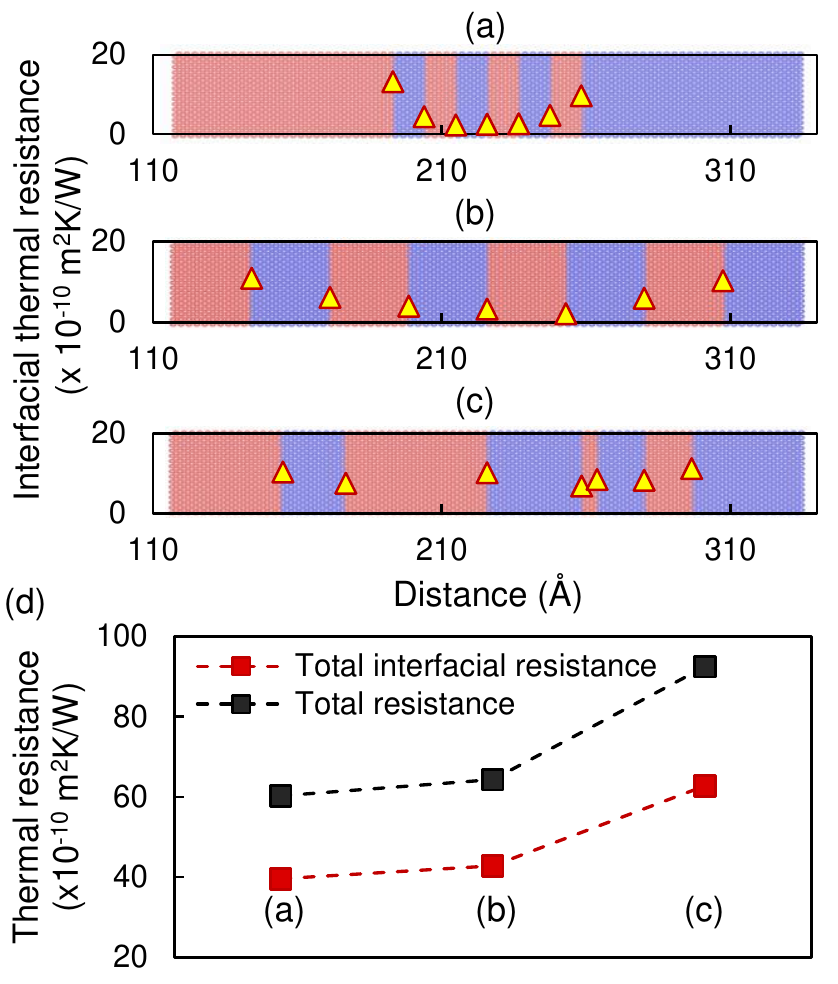}
\caption{\label{fig:res} Calculated thermal resistances at all interfaces (yellow triangles) in three different 40 UC RML structures: (a) RML with high $\kappa$ identified by the ML enabled search (b) 5-5 SL (c) RML with low $\kappa$ identified by manual random search. The RML structures are underlaid for ease of visualization. (d) Comparison of the total interfacial thermal resistance (black squares) and total thermal resistance (red squares) for the three RML structures (a-c).}
\end{figure}

Finally, the identified exceptional RML structures shown in Fig.~\ref{fig:NNresults} (d) are studied to understand the underlying phonon transport characteristics leading to the disorder induced enhancement of $\kappa_l$. We observe the presence of small layer thicknesses due to closely spaced interfaces in these structures, which we attribute as the cause for the increased thermal transport. At an SL period of $5.4$ nm, the relatively large layer thicknesses are above the coherence length of most phonons, as a result of which the contribution of coherent phonon transport to the SL $\kappa_l$ is quite low. However, the reduced thicknesses of some layers in the identified RMLs lead to an increased coherent phonon contribution, whereby the apparent thermal resistance of the interfaces are lowered. To verify our hypothesis, we calculated the total resistance across the RML as well as the contribution of the apparent interface resistances for three different 40 UC structures: $(i)$ the RML with $\kappa_l$ higher than the $5-5$ SL identified through our search process, $(ii)$ the $5-5$ SL and $(iii)$ a RML with low $\kappa_l$ identified by the random search. The apparent interfacial thermal resistances at each of the interfaces in the RML structure are shown in Fig.~\ref{fig:res} (a-c), superimposed on the visual representation of the RML structure. We can see that the compared to the $5-5$ SL (Fig.~\ref{fig:res} (b)), the apparent interfacial resistances are visibly reduced in the high $\kappa$ RML (Fig.~\ref{fig:res} (a)), which is the effect of a higher amount of coherent phonon transport. As a result, the RML shows a lower total interfacial thermal resistance and total thermal resistance than the SL, as seen in Fig.~\ref{fig:res} (d). Finally, the localization of coherent phonon modes due to sufficient layer thickness randomization in the RML structure shown in Fig.~\ref{fig:res} (c) and the absence of many closely spaced interfaces leads to a higher interfacial resistance and lower $\kappa_l$, which is in accordance to the previously accepted hypothesis. Our results indicate that randomness of layer thicknesses in SLs can be engineered to have dual effects via tuning the contribution of coherent phonons, which can either decrease or enhance thermal conductivities. Generally, in short period SLs, randomness can cause localization of coherent phonons and reduce $\kappa_l$. On the other hand, certain forms of aperiodicity in large period SLs can enable stronger coherent phonon transport that is not localized, thus enhancing $\kappa_l$.

\section{Conclusions}

In summary, we demonstrate an iterative machine learning approach for discovering exceptional thermal transport physics. Although it is generally accepted that randomization of layer thicknesses of a binary periodic superlattice lowers its cross-plane $\kappa_l$, we aim to find structures showing the opposite trend, \textit{i.e.} an enhancement of $\kappa_l$ due to disorder. We employ a convolutional neural network to rapidly predict the thermal conductivities of all RMLs in the design space. The training dataset is generated in an iterative method in order to help the CNN dynamically learn the spatial features leading to locally enhanced phonon transmission. Our CNN accelerated search is able to identify RML structures with higher $\kappa_l$ than the superlattice at an average period of $5.4$ nm, which is attributed to an increase in coherent phonon contribution and decrease in apparent interfacial thermal resistance at closely spaced RML interfaces as compared to the SL. Our results demonstrate the ability of machine learning based methods to help discover exceptions and low-probability-of-occurrence solutions in a large search space.

\section{Acknowledgements}

This work was supported by the Defence Advanced Research Projects Agency (Award No. HR0011-15-2-0037) and the School of Mechanical Engineering, Purdue University. Simulations have been performed at the Rosen Center of Advanced Computing at Purdue University.

\section{Vitae}

\begin{wrapfigure}{l}{30mm} 
    \includegraphics[height=1.25in,width=1in,clip,keepaspectratio]{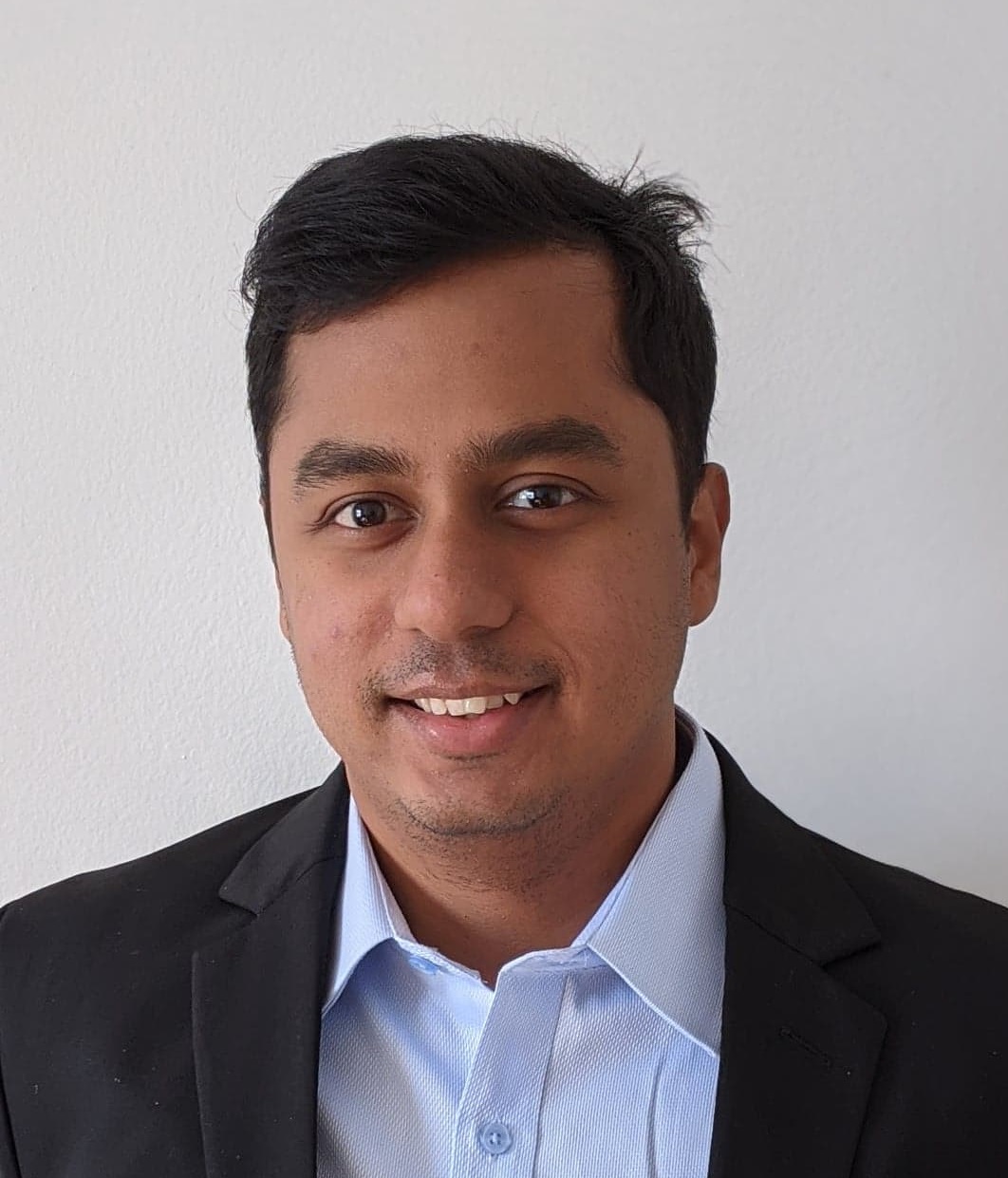}
\end{wrapfigure}
\textbf{Prabudhya Roy Chowdhury} 

Prabudhya is a Ph.D. candidate at Purdue University, and his current research interest is investigating thermal transport in nanostructures using a combination of atomistic simulations and machine learning techniques. He received his B.S. and M.S. from the Indian Institute of Technology, Kharagpur in 2016. He is the recipient of the Bilsland Dissertation Fellowship at Purdue University, and was a summer research intern at the Assembly, Test and Technology Development Division at Intel Corporation.

\begin{wrapfigure}{l}{30mm} 
    \includegraphics[height=1.25in,width=1in,clip,keepaspectratio]{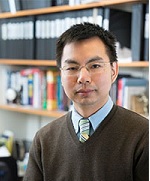}
\end{wrapfigure}
\textbf{Xiulin Ruan} 

Dr. Xiulin Ruan is a professor in the School of Mechanical Engineering and the Birck Nanotechnology Center at Purdue University. He received his B.S. and M.S. in Engineering Thermophysics from Tsinghua University in 2000 and 2002, respectively. He received an M.S. in Electrical Engineering and Ph.D. in Mechanical Engineering from the University of Michigan at Ann Arbor, in 2006 and 2007 respectively. His research interests are in multiscale multiphysics simulations and experiments of phonon, electron, and photon transport and interactions, for various emerging applications.


\providecommand{\noopsort}[1]{}\providecommand{\singleletter}[1]{#1}%

\end{document}